\RequirePackage{lineno}

\documentclass[12pt]{article}
\usepackage{graphicx}
\usepackage{lineno}

\def\nova{NO$\nu$A}

\newcommand\pubnumber{DPF2015-244}
\newcommand\pubdate{\today}

\def\napoli{Argonne National Laboratory, HEP Division, 9700 S Cass Ave,  Lemont, IL 60439}

\def\Title#1{\begin{center} {\Large #1 } \end{center}}
\def\Author#1{\begin{center}{ \sc #1} \end{center}}
\def\Address#1{\begin{center}{ \it #1} \end{center}}

\newcommand\pubblock{\rightline{\begin{tabular}{l} \pubnumber\\
         \pubdate  \end{tabular}}}
\newenvironment{Abstract}{\begin{quotation}  }{\end{quotation}}
\newenvironment{Presented}{\begin{quotation} \begin{center} 
             PRESENTED AT\end{center}\bigskip 
      \begin{center}\begin{large}}{\end{large}\end{center} \end{quotation}}

\def\beq{\begin{equation}}
\def\eeq#1{\label{#1}\end{equation}}
\def\eeqn{\end{equation}}

\def\beqa{\begin{eqnarray}}
\def\eeqa#1{\label{#1}\end{eqnarray}}
\def\eeqan{\end{eqnarray}}

\let\bar=\overbar

\def\Dslash{\not{\hbox{\kern-4pt $D$}}}
\def\dslash{\not{\hbox{\kern-2pt $\del$}}}

\def\msb{{\bar{\ssstyle M \kern -1pt S}}}

\begin{document}
\begin{titlepage}
\pubblock

\vfill
\Title{Extrapolation Techniques and Systematic Uncertainties in the NO$\nu$A Muon Neutrino Disappearance Analysis}
\vfill
\Author{Louise Suter, for the  \nova~Collaboration}
\Address{\napoli}
\vfill
\begin{Abstract}
The NOvA long-baseline neutrino experiment consists of two highly active, finely segmented, liquid scintillator detectors located 14 mrad off Fermilab's NuMI beam axis, with a Near Detector located at Fermilab, and a Far Detector located 810 km from the target at Ash River, MI.
NO$\nu$A  released it first preliminary results of the muon neutrino disappearance parameters, measuring $\sin^2(\theta_{23}) =~0.51~\pm~0.10$ and or the normal hierarchy $\Delta~m^2_{32}~=~2.37^{+0.16}_{-0.15}~\times~10^{-3}$~eV$^2$ and for the inverted hierarchy
$\Delta m^2_{32}~ =~-2.40^{+0.14}_{-0.17}~\times~10^{-3}$~eV$^2$. 
 This talk will present a discussion of the systematic uncertainties and extrapolation methods used 
 for this first analysis which uses $2.74\times10^{20}$ POT-equivalent collected between July 2013 and March 2015. 

\end{Abstract}
\vfill
\begin{Presented}
DPF 2015\\
The Meeting of the American Physical Society\\
Division of Particles and Fields\\
Ann Arbor, Michigan, August 4--8, 2015\\
\end{Presented}
\vfill
\end{titlepage}
\def\thefootnote{\fnsymbol{footnote}}
\setcounter{footnote}{0}

\section{Introduction}

\begin{figure}[h]
\begin{minipage}{18pc}
\includegraphics[width=16pc]{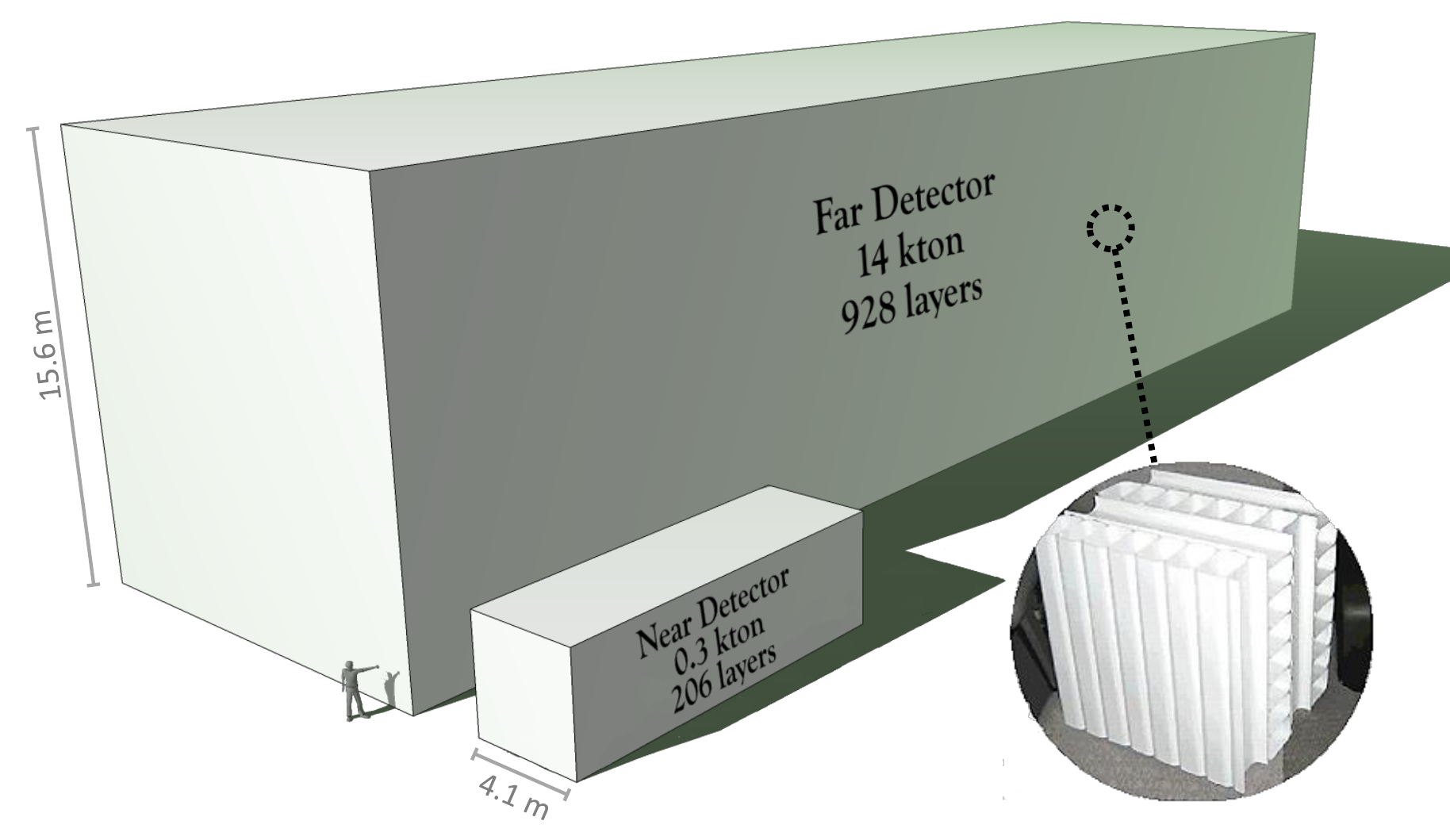}
\caption{\label{nova2} The relative sizes of the \nova~Far and Near Detectors. The structure of the \nova~detector layers is also shown.}
\end{minipage}\hspace{2pc}%
\begin{minipage}{18pc}
\includegraphics[width=16pc]{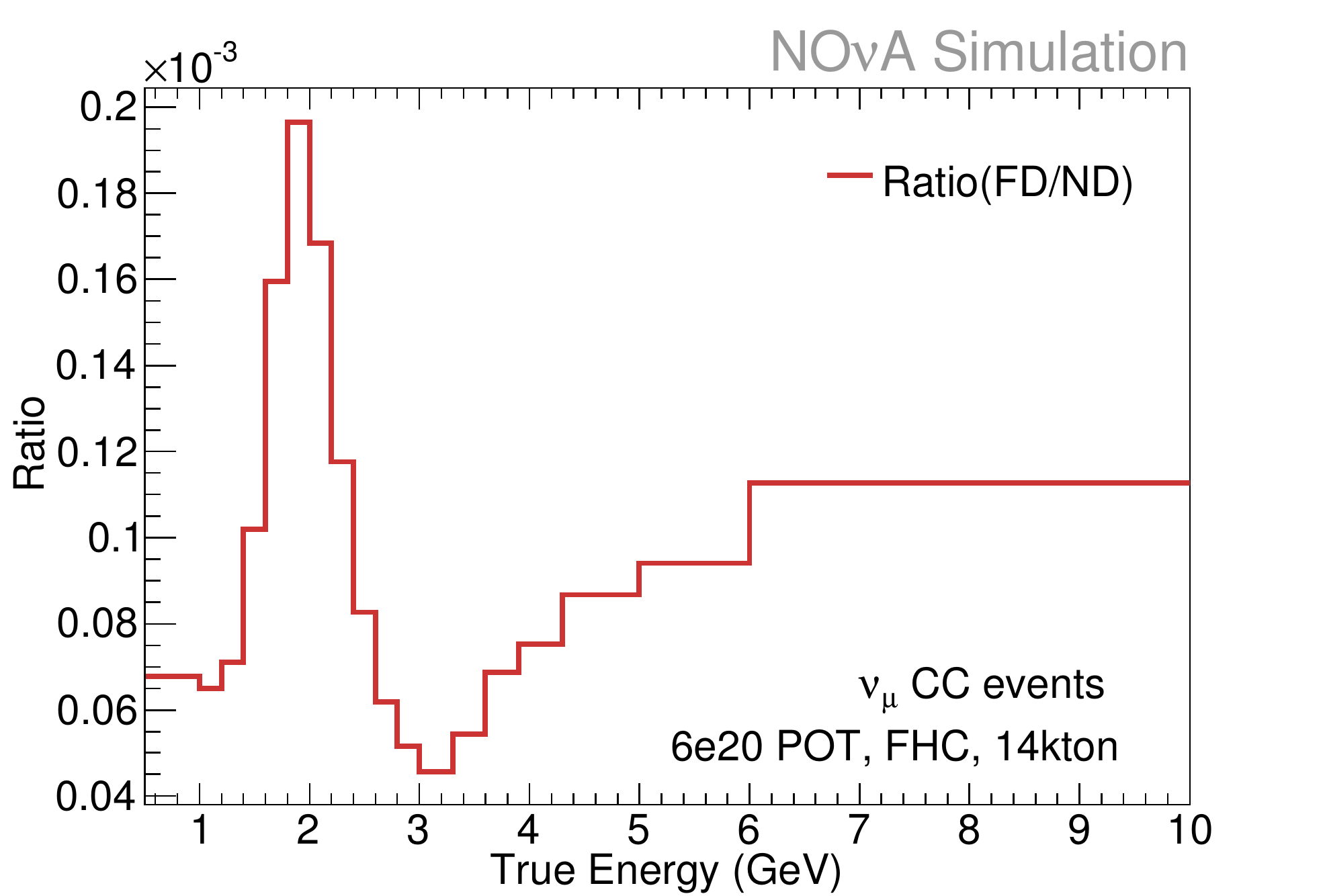}\hspace{2pc}%
\caption{\label{nu-e} The simulated ratio of the Far to the Near Detector flux for charged current $\nu_\mu$ events as a function of true energy. }
\end{minipage} 
\end{figure}

The \nova~experiment, a long baseline neutrino oscillation experiment, 
consists of two almost completely active, segmented, liquid scintillator detectors. The 0.3~kton Near Detector (ND) is located on site at Fermilab, 105 m underground and 1 km away from Fermilab's NuMI beam production target. The 14 kton Far Detector (FD) is located at Ash River, Minnesota, 810 km away from Fermilab's NuMI neutrino source.
The FD building is covered by a 3 m equivalent mound of barite rock. This provides an overburden of more than ten radiation lengths to reduce background from cosmic rays.
The relative sizes of the detectors are shown diagrammatically in Figure~\ref{nova2}.

The two detectors are located 14.6 mrad off the NuMI beam axis, resulting in a relatively narrow neutrino energy band centered at 2 GeV, where
 the $\nu_\mu$($\bar{\nu}_{\mu}) \rightarrow  \nu_e(\bar{\nu}_{e})$ oscillation maximum occurs (see Figure~\ref{flux}). This narrow band beam results not only in a increased flux of events at 2 GeV events but also in a suppression of the neutral current background which is very important for $\nu_\mu$ to $\nu_e$ measurements. 
 The neutrino energy relies on the angle between pion decay and neutrino interaction inside the detector. As one goes to an off-axis location the dependence on pion energy becomes flat. 
 
Both \nova~detectors are highly segmented tracking calorimeters built in their entirety from low Z (0.18 radiation lengths per layer) and highly reflective (15\% TiO2) PVC cells~\cite{TDR}. The PVC cells are filled with liquid scintillator consisting of mineral oil infused with 5\% pseudocumene.
The detectors are constructed from extrusions consisting of planes of 6 cm $\times$ 4 cm cells, where each cell extends the full width or height of the detector. These PVC extrusions are assembled in alternating layers either vertically or horizontally, as can be seen in Figure~\ref{nova2}. This orientation of the cells allows for 3D event reconstruction.
In total there are  344,054 cells in the FD and 21,192 cells in the ND. 
The scintillation light is collected in every cell by a loop of wavelength shifting fiber and each cell is read out individually, using 32-pixel avalanche photo-diodes (APDs).

The NO$\nu$A FD has been taking data since July 2013, taking advantage of the modular nature of NO$\nu$A. The first data was recorded with the first 1 kton block of the detector, with additional kton blocks being added once they were fully commissioned. This allowed for the detector to take data as it was constructed. 
Both NO$\nu$A detectors have been fully constructed and commissioned since August 2014.  As the detector volume was changing size as the early data was recorded this information is encoded in POT quoted, hence exposure is giving the POT-14-ton-equivalent. 

The first results from the NO$\nu$A experiment were presented in August 2015 
and these proceedings will present a discussion of the systematic uncertainties associated 
 with the analysis of first $2.74\times10^{20}$ POT-equivalent collected for the muon-neutrino disappearance analysis. This data was collected  between July 2013 and March 2015. 
The methods developed to predict the FD spectrum as extrapolated from the observed ND data will also be presented.

\section{Near to Far Detector Extrapolation Method}

The neutrino energy spectrum at the \nova~ND is measured close to the neutrino source before neutrino oscillations have occurred. This large statistics data sample is used to validate the Monte Carlo (MC) prediction of the expected beam flux and the simulation of the detector response.  All beam intrinsic backgrounds can be measured at the ND to a high precision. 

\nova~uses the ND energy spectrum to make a prediction of the energy spectrum that will  be seen at the \nova~FD.  This prediction technique is known as extrapolation and reduces the dependence on systematics uncertainties which apply to both detectors. 
\nova~employs a direct extrapolation technique where the ratio of the FD to ND flux, as determined from MC, was used to predict the expected FD energy spectrum from the measured ND energy spectrum. This extrapolation is performed bin-by-bin in reconstructed energy. To extend the extrapolation technique beyond the prediction of the null oscillation spectrum,
migration matrixes are used to apply  corrections to the measured ND reconstructed energy spectra to obtain the MC true energy spectra. This allows for oscillation probability predictions to be applied. 
To fully qualify the oscillation parameters that describe the observed FD spectrum we minimize $\chi^2$ between observed FD data best-fit, fitted using the full systematic suite (described in Section~\ref{sys}), and the predicted FD spectrum under different oscillation predictions. 
The full three-flavor parameterization of neutrino oscillations is used, with the other oscillation parameters and their uncertainties marginalized over.  The oscillation parameters included in fit  are;  $\Delta m^2_{21} = 7.53 \pm 0.18\times 10^5 $eV$^2$;   $\sin^2(2\theta_{13}) = 0.086 \pm 0.005$; $\sin^2(2\theta_{12}) = 0.846 \pm 0.021 $; and $\delta_{CP} $ is unconstrained.

\begin{figure}[h]
\begin{center}
\includegraphics[width=17.5pc]{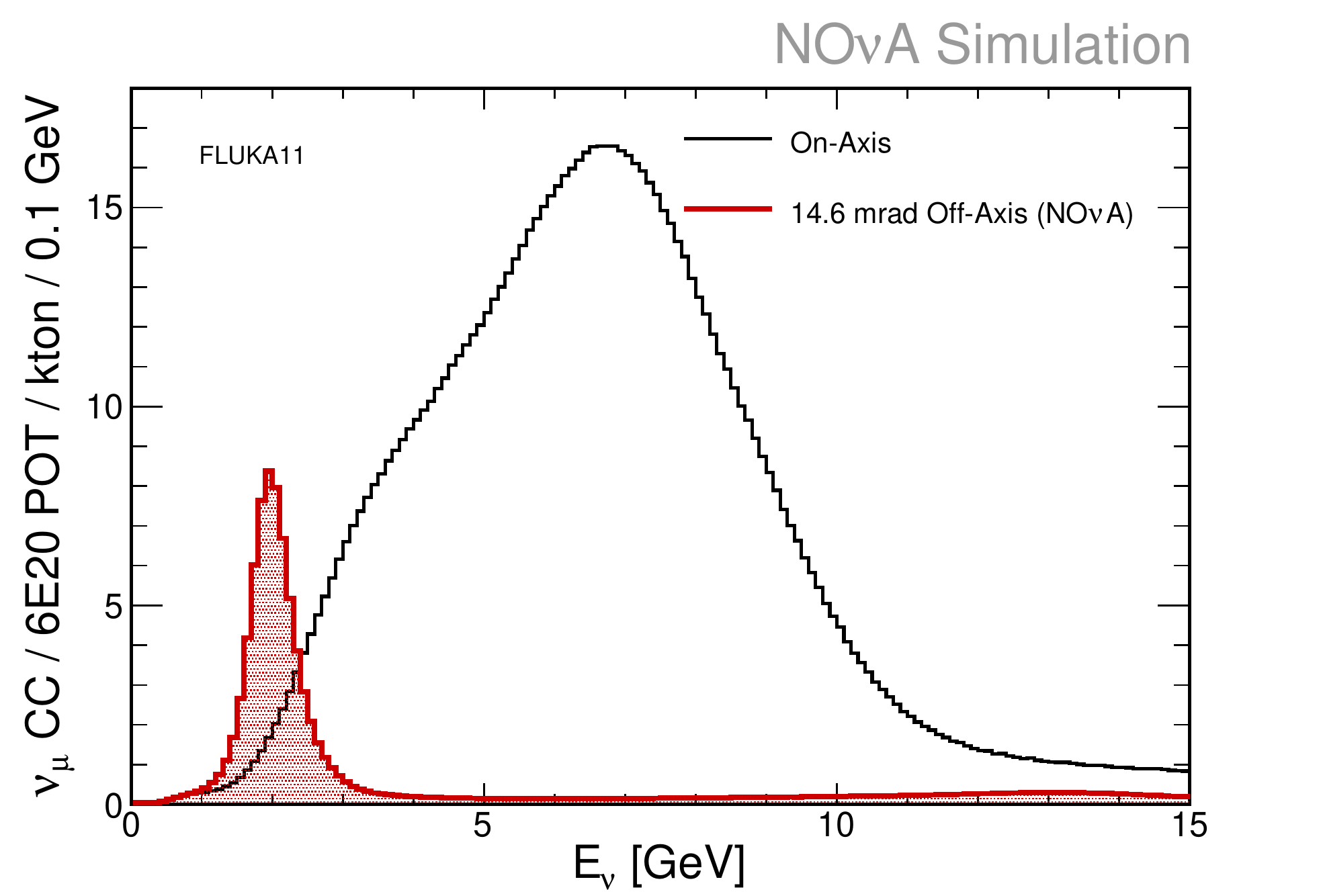}
\includegraphics[width=17.5pc]{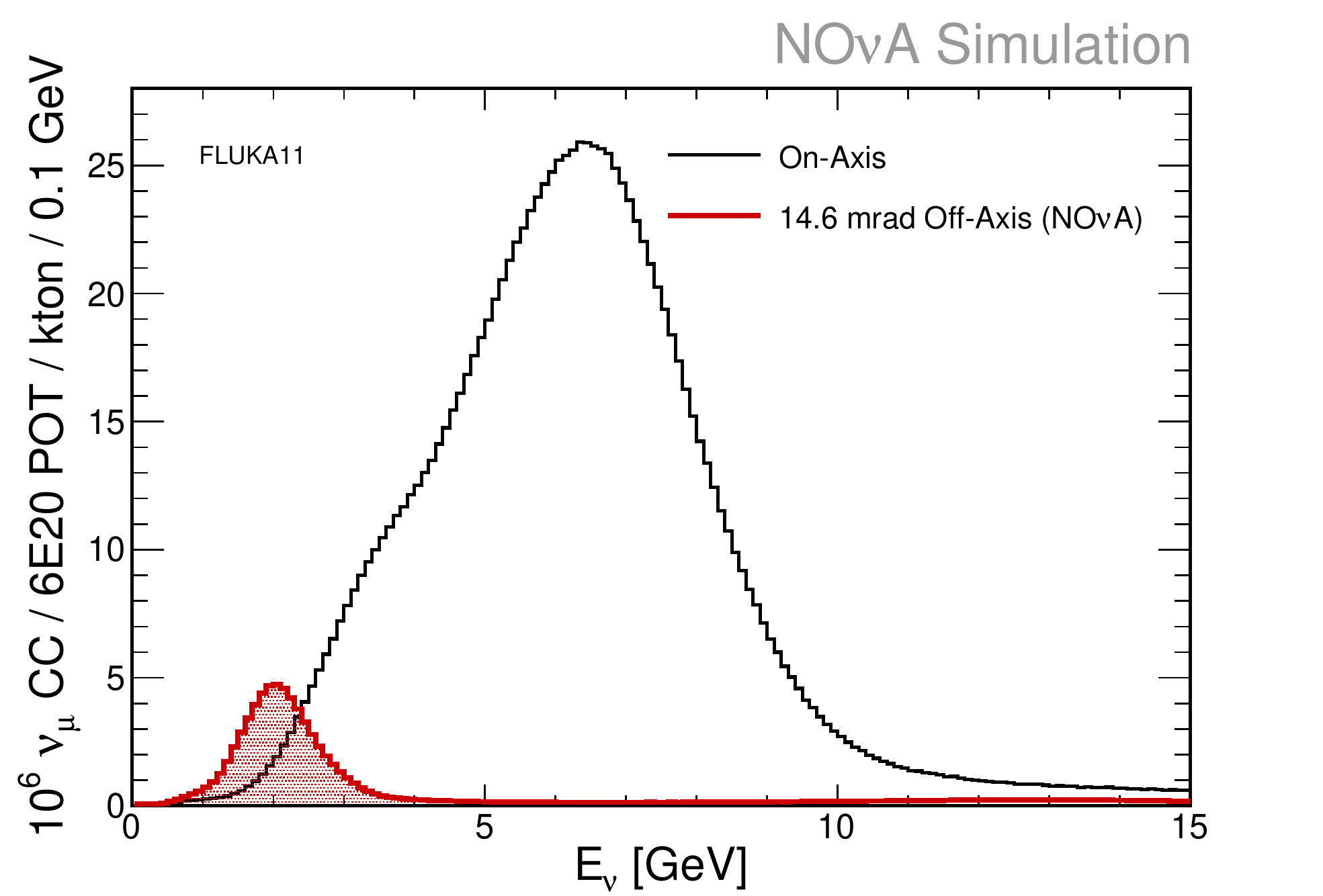}
\end{center}
\caption{\label{flux} The energy spectrum for $\nu_{\mu}$ charged current events both on-axis (open histogram) and 14.6 mrad off-axis (red histogram), in the NuMI beam. The left spectrum is for the \nova~FD and the right is for the ND.}
\end{figure}

As the two detectors are functionally identical this ratio based method allows for reductions in detector-response, object-identification, and energy reconstruction based uncertainties. Slight differences in acceptance, due to the size of the detector, and in the flux lead to not complete cancellation of systematics uncertainties. 
The MC flux prediction results in one of the largest single detector systematic uncertainties. 
The two detectors see slightly different  fluxes so this uncertainty does not cancel completely. 
This arrises as the ND sees a line source around 14.6 mrad where as the FD a point source at exactly 14.6 mrad, see Figure~\ref{flux}.

The \nova~experiment performs a blinded analysis technique where are all tools and algorithms are constructed using simulations or side band regions, only once an analysis is classified, by the collaboration, to be complete is the analysis run over the FD data.

\section{Systematic Uncertainties}
\label{sys}

I will discuss the non-negligible systematic uncertainties associated with the NO$\nu$A muon neutrino disappearance analysis. Multiple other effects where considered, for example the detector response modeling and the attenuation calibration corrections, but will not be discussed here as they were determined to be negligible. A summary of all the non-negligible systematic uncertainties is given in Table~\ref{tab}.\\

\noindent \textbf{a) Uncertainty of Background Rates}\\
The only non-negligible contaminations in the in the selected muon-neutrino sample (backgrounds) for the muon-neutrino disappearance analysis are the neutral current and tau neutrino backgrounds. These contamination rates are estimated from simulation and a 100\% uncertainty is taken on them. 

The largest background at the NO$\nu$A FD is the 
rate of cosmic muons. This rate is determined from minimum-bias data taking outside of the neutrino beam spill. 
The statistical uncertainty of this minimum-bias sample is negligible, as along with each beam spill a much larger (35x) minimum-bias sample  is recorded. This sample is recorded using the same detector conditions so they can be directly matched. \\

\noindent \textbf{b) Calibration uncertainty: Absolute Hadronic Energy Scale} \\
The NO$\nu$A experiment uses muons which stop in the detector to provide a standard candle for setting the absolute energy scale. The 
uncertainty on this is estimated from maximum difference between the multiple probes of calibration which are available at NO$\nu$A. The observed difference is propagated through the full analysis framework, including the extrapolation and oscillation parameter minimization. The probes available at NO$\nu$A include the Michele electrum spectrum, the $\pi^0$ mass peak and the $dE/dx$ of the muon and the proton. Using this method a 5\% percent absolute and a 5\% relative calibration uncertainty are determined. \\

Comparing the the off-track energy measured in NO$\nu$A ND charged current muon-neutrino interactions to the simulation a discrepancy is seen. We define off-track  energy to be the sum of all energy associated with the neutrino interaction that is not part of the muon track.  This is referred to as Hadronic Energy. 
As the NO$\nu$A detectors are located off-axis the location of the neutrino energy peak at this location is known to a high precision. Therefore we use this knowledge to tune the hadronic energy such at the neutrino energy is peaked, as expected, at 2 GeV. 
 Using the ND data a 21\% hadronic energy correction is determined. This correction translates into a 6\% correction to the neutrino energy. 
We conservatively take a 100\% absolute uncertainty on this correction. This is our largest systematic uncertainty.
Combining this correction with the absolute hadronic energy scale we get a 22\% total absolute hadronic energy uncertainty.  \\

\begin{figure}[h]
\begin{center}
\includegraphics[width=17.5pc]{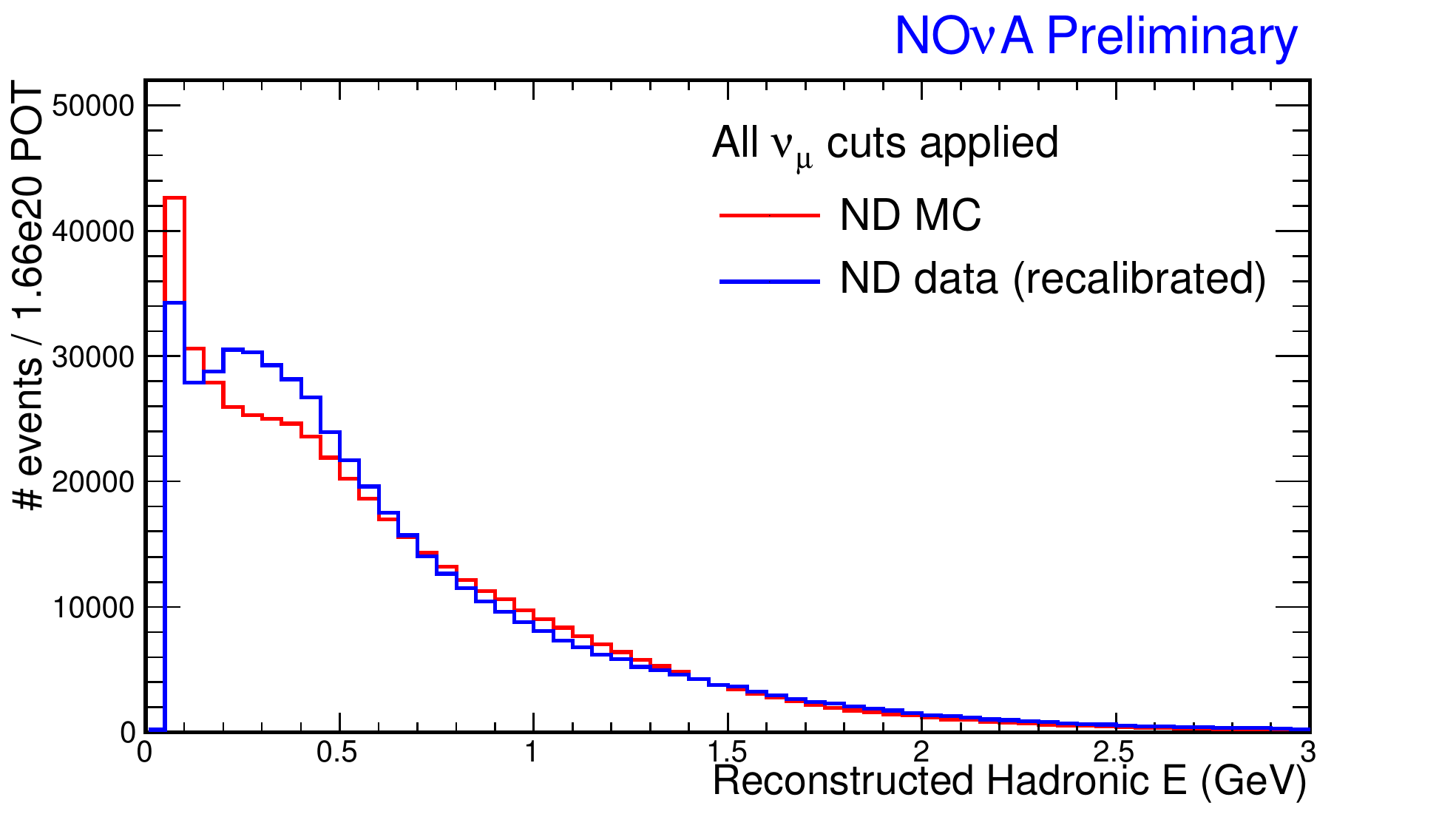}
\includegraphics[width=17.5pc]{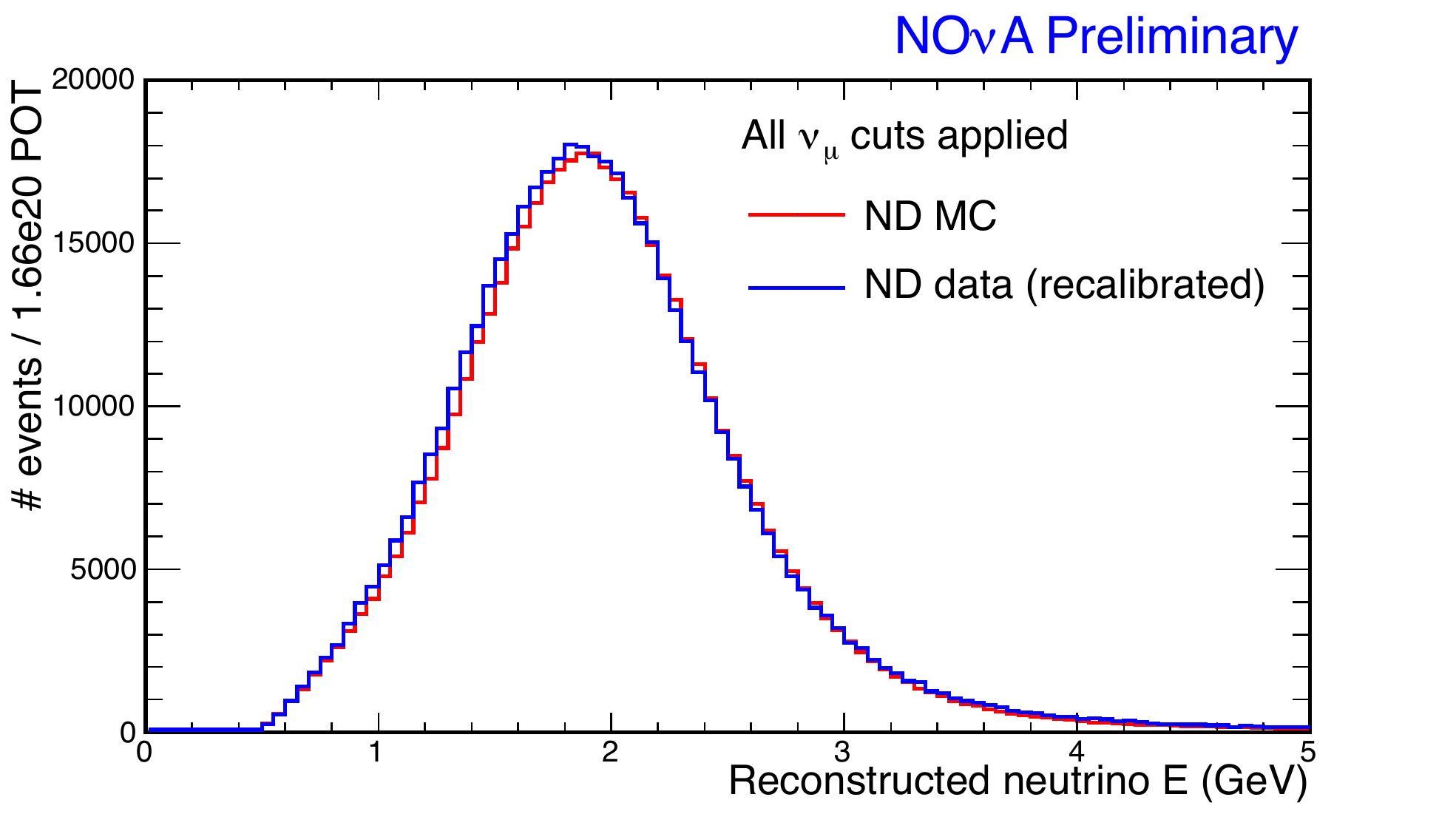}
\end{center}
\caption{\label{hadE} The distribution of the off-track energy (left) and the reconstructed neutrino energy (right) shown for both the simulated ND events (red) and the recalibrated ND data (blue) after the 21\% correction factor is applied.  }
\end{figure}

\noindent \textbf{c) Calibration uncertainty: Relative Hadronic Energy Scale}\\
In addition, we calculate the relative hadronic energy uncertainty due to the different detector acceptances. 
As the 21\% correction factor is calculated using ND data it may be optimized only for the ND. Due to the smaller size of the ND the acceptance is sculpted as compared to the FD and a higher percentage of the events that pass the selection are quasi-elastic. 
 This effect is
investigated by allowing the normalization and the energy scale of deep inelastic scattering, resonant and quasi-elastic events (as defined by GENIE~\cite{genie}) to float. A 
 three parameter simultaneous fit of the muon energy, the off-track energy and normalization is done. The difference between the 
 one-parameter 21\% scaling and this interaction-dependent scaling is used to determine the relative uncertainty.
A 2\% relative uncertainty and 1\% relative normalization uncertainty  are determined. The relative uncertainty  is combined with the uncertainty discussed in b) to give a  5\% total relative hadronic energy uncertainty. The  distribution of the off-track energy  and the reconstructed neutrino energy  at the NO$\nu$A ND are shown in Figure~\ref{hadE}.  \\

\begin{figure}[h]
\begin{center}
\includegraphics[width=11.8pc]{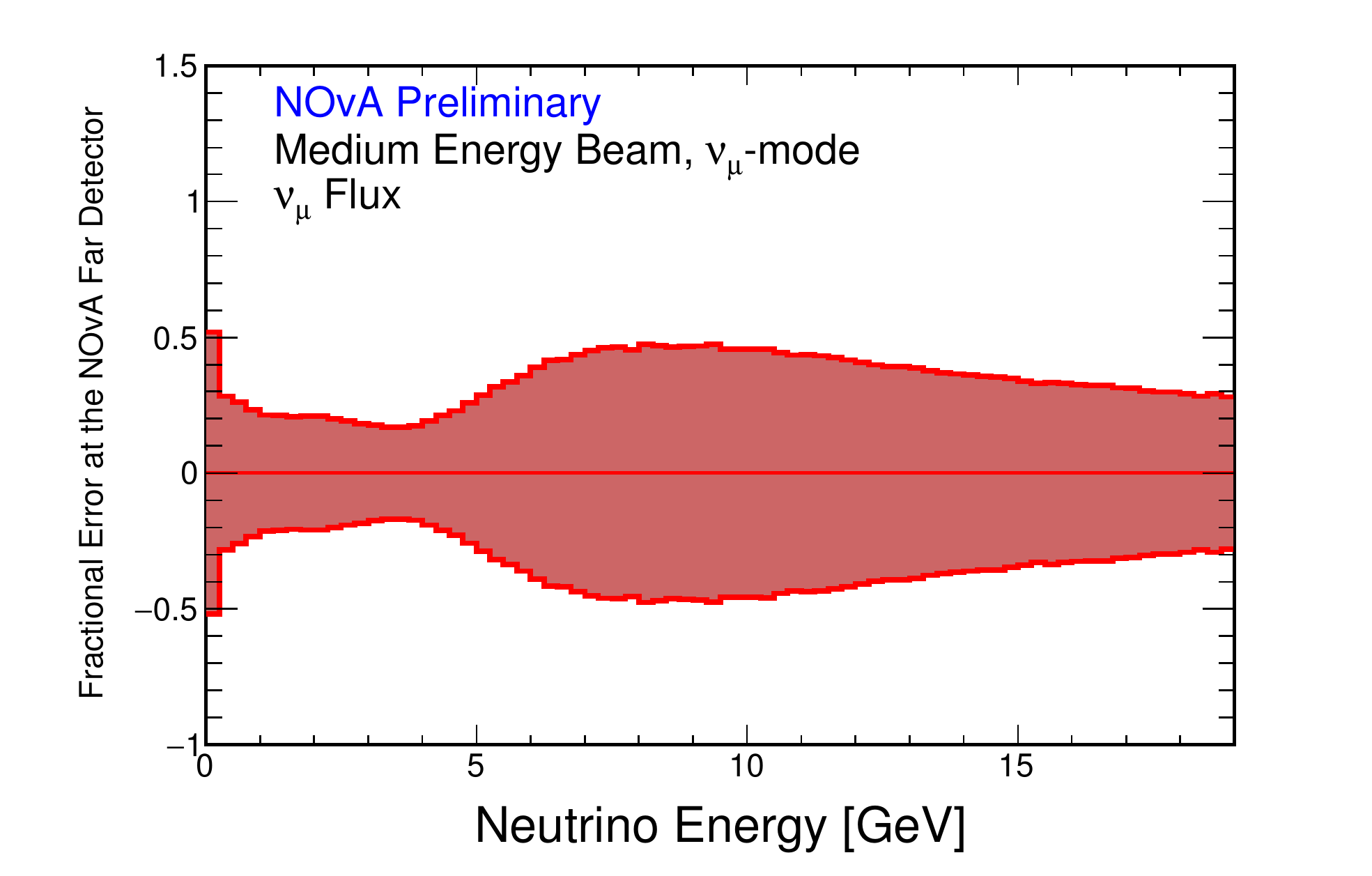}
\includegraphics[width=11.8pc]{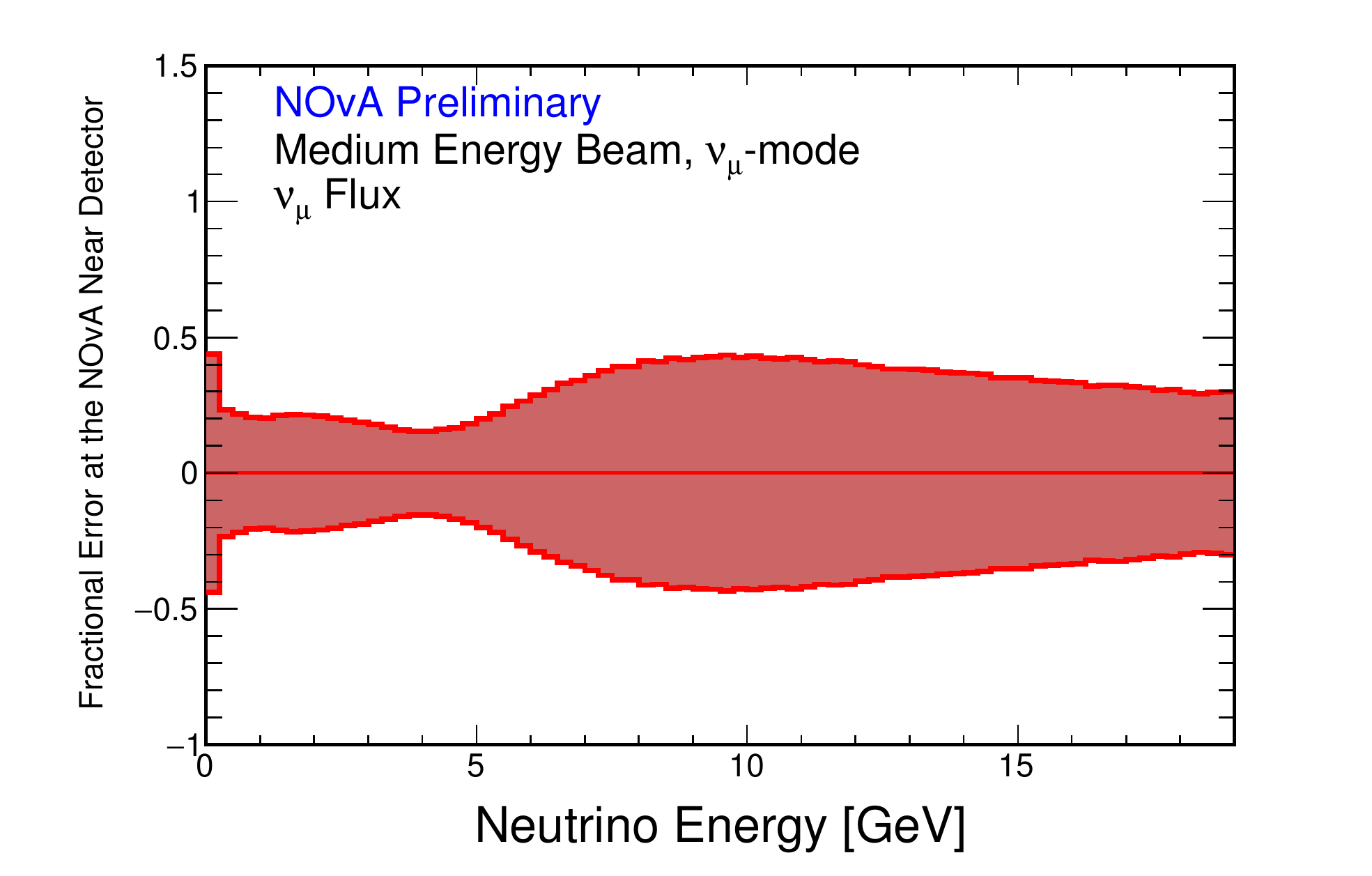}
\includegraphics[width=11.8pc]{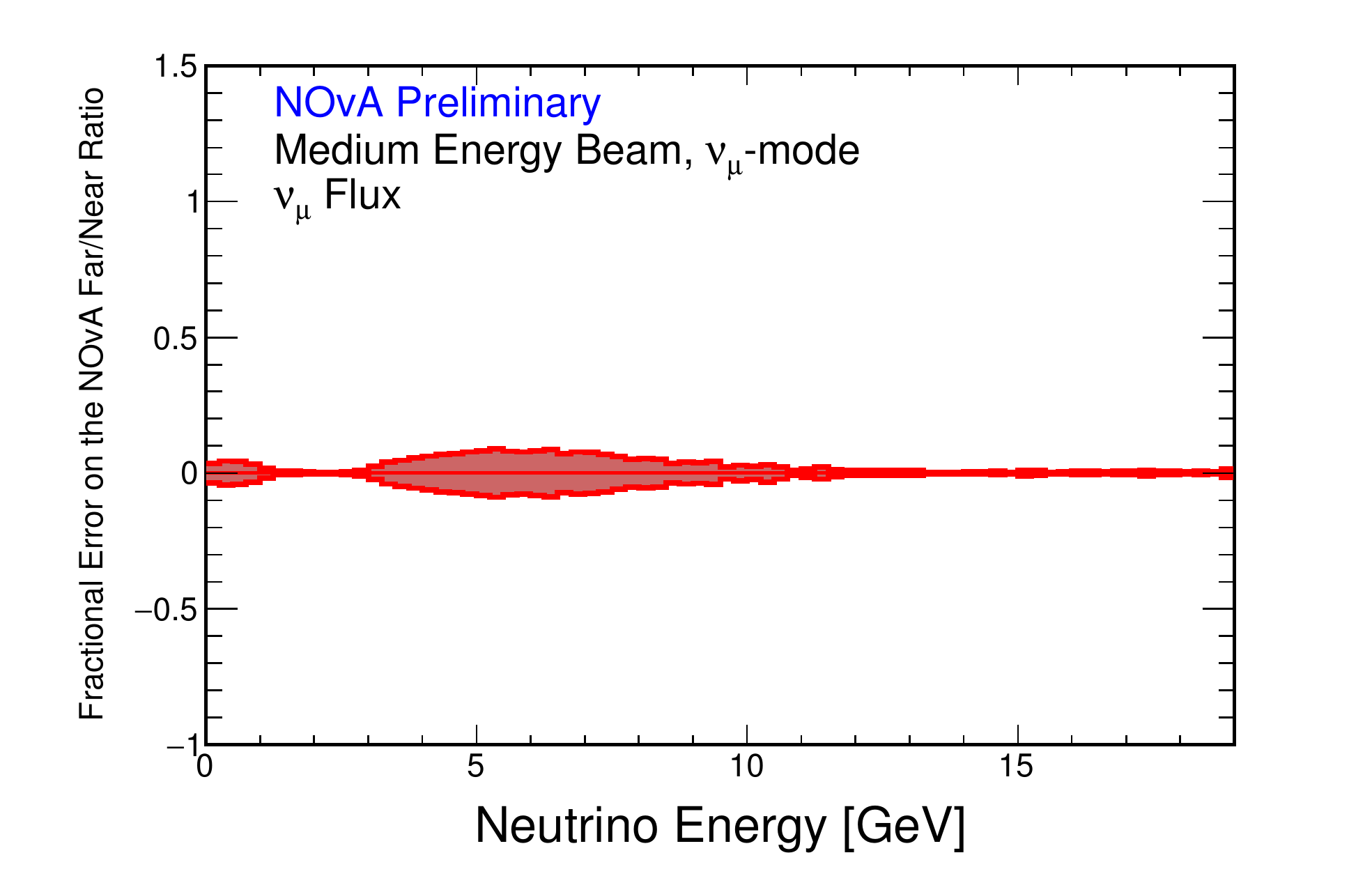}
\end{center}
\caption{\label{fluxu} The NO$\nu$A flux uncertainty for the far detector (left), near detector (middle) and the ratio (right). }
\end{figure}

\noindent \textbf{d) Flux Uncertainties} \\
The NO$\nu$A flux is modeled using FLUKA/FLUGG~\cite{fluka}. 
For each individual detector the flux uncertainty is large (20\% at the 2 GeV peak) and dominated by the hadron production uncertainties. The hadron production uncertainties are 
estimated by comparing the NuMI target MC predictions to the the thin-target data from NA49~\cite{na49}. 
The hadron transport uncertainties were also investigated. Uncertainties due to the 
NuMI target and horn positions, the horn current and the magnetic field, and the beam spot size and position were 
determined to be small compared to hadron production uncertainties and are considered negligible. 
The flux uncertainties are highly correlated between the two detectors. For each individual detector the flux uncertainty is large but  due to the use of the extrapolation method it is the ratio of the uncertainties that is relevant. As the fluxes are very highly correlated between the two detectors  the flux uncertainty is reduced to the percent level. The fraction uncertainty on the NO$\nu$A ND and FD and the ratio is shown in Figure~\ref{fluxu}. \\

\noindent \textbf{e) Absolute Normalization}\\ 
There are two sources of absolute normalization uncertainty on NO$\nu$A. The first arrises from the 
an uncertainty in the detector mass which leads to uncertainty in the exposure. The NO$\nu$A detectors are 
constructed from PVC cells which are filled with a liquid scintillator. Each cell contains a loop of wavelength shifting fiber. These cells are extruded in sets of 16 which are glued together to make a plane of the detector. A
0.7\% normalization uncertainty is taken on the amount of plastic, glue, scintillator and wavelength shifting fiber. This is determined from the uncertainties on these components as built and as compared to what is in the simulation. 
The second source of absolute normalization is due to a potential proton-on-target, POT,   skew between the two detectors. 
As data taking at the ND and FD  was over different periods if there had been a POT  mis-measurement this could result in a normalization skew. The NuMI beam has been shown to be very stable and a conservative 0.5\% proton-on-target normalization uncertainty is taken. 
Combining this with the mass uncertainty gives an overall 0.9\% normalization uncertainty\\

\noindent \textbf{f) Neutrino Interaction Modeling}\\
NO$\nu$A uses GENIE to study the uncertainty on cross sections and final state particles exiting the nucleus.  The effect of 1 and 2 $\sigma$ variations of the 67 parameters provided in GENIE on the muon-neutrino charged-current energy spectrum was studied.
Of these 67 parameters only 6 were seen to have a noticeable effect. There are; 
the axial mass of the charged current quasi-elastic cross section; the axial mass of the neutral current quasi-elastic cross section; the axial mass of the charged current resonant cross section; the axial mass of the neutral current resonant cross section; the 
 vector mass for the charged current resonant cross section; and the vector mass for the neutral current resonant cross section. 
As well as these 6 largest, and an effective parameter that includes the effect of the other 61 parameters added in quadrature, was added as penalty terms in the fit. From this a 10 - 25\% uncertainty on neutrino interaction dynamics was determined but again this uncertainty mostly cancels out due to the use of a ratio method.

\begin{table}[t]
\begin{center}
\begin{tabular}{l|cc}  
Systematic & Value (1$\sigma$) & Best fit ($\sigma$)\\
\hline
Bkg. (neutral current and $\nu_\tau$) &  100\% &  0.06\\
Absolute Normalization & 1.3\% & 0.0008\\
Absolute Hadronic energy scale & 22\%  & -0.67\\
Absolute energy scale & 1\% & 0.06\\
Beam  & Energy dependent  & -0.02\\ 
& (20\% at 2 GeV) & \\ 
Relative Normalization &  1.4\% & -0.03\\
Relative Hadronic energy scale &5.4\% &0.05\\
GENIE $M_a$ &15-25\% &-0.18\\
GENIE $M_v$ &10\% &-0.06\\
\end{tabular}
\caption{Summary table of the non-negligible systematics in NO$\nu$A muon neutrino oscillation measurement.}
\label{tab}
\end{center}
\end{table}

\section{Results}

\begin{figure}[h]
\begin{center}
\includegraphics[width=17.5pc]{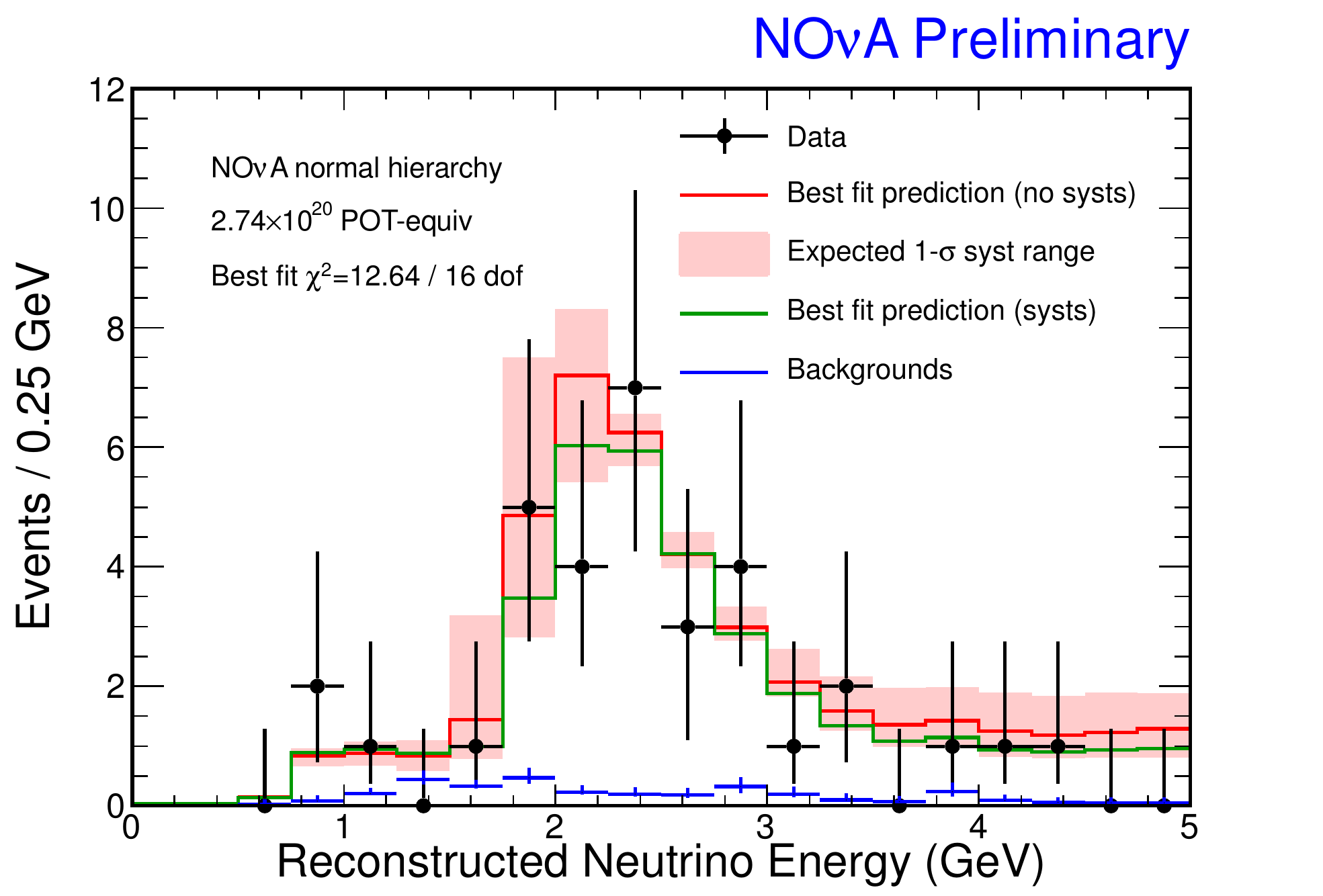}
\includegraphics[width=17.5pc]{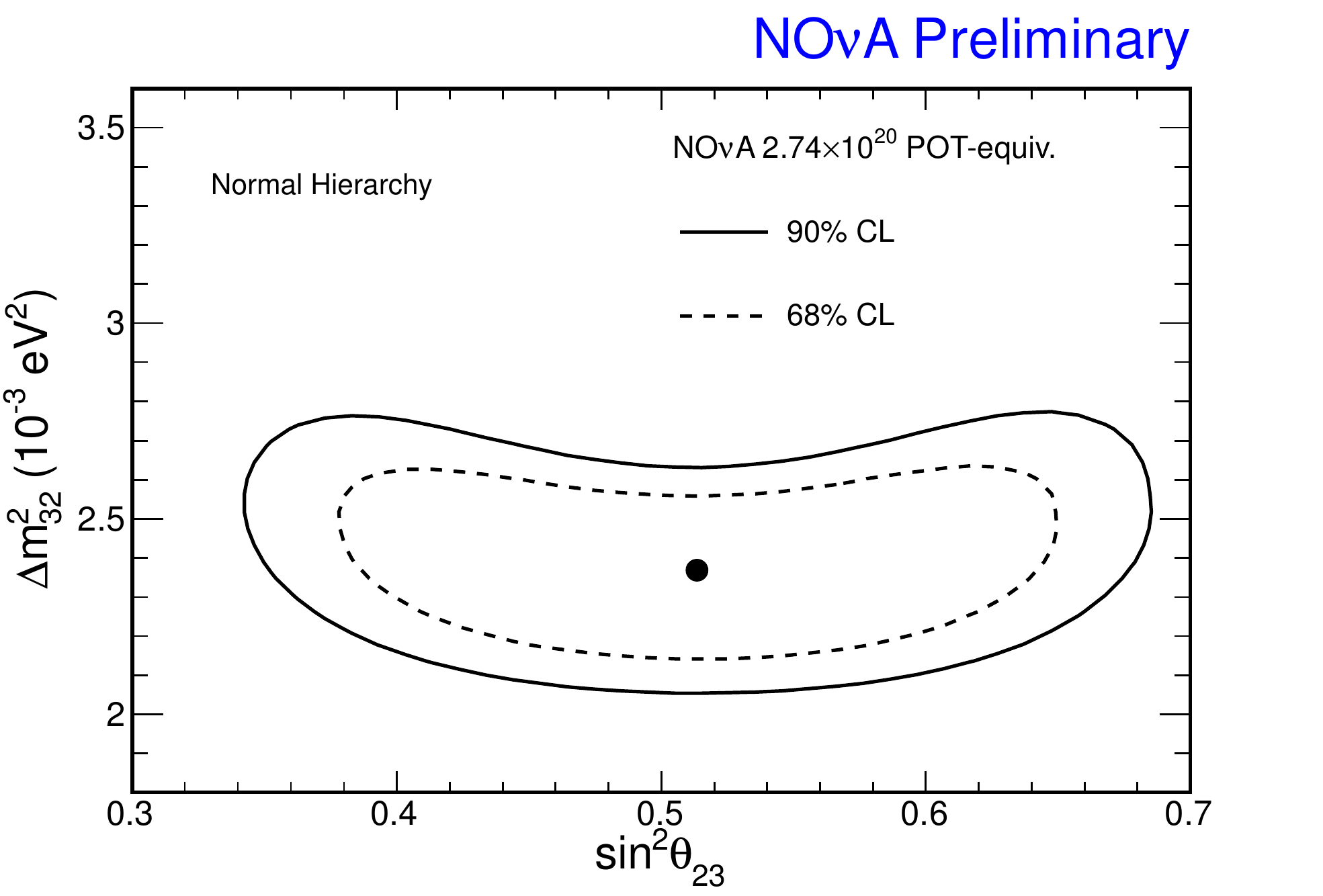}
\end{center}
\caption{\label{results} The energy spectrum of the observed events, along with the best fit distribution to these events (with and without systematics) (right). The best fit point along with the 68\% and 90\% contours in $\sin^2(\theta_{23})-\Delta m^2_{32}$ space for the normal hierarchy (left).}
\end{figure}

NO$\nu$A predicted a event rate of 201 $\nu_\mu$ charge current events at its FD extrapolated from the ND data, in range 0 -- 5 GeV. This included a predicted background of $1.4 \pm 0.2$ comic muons determined from minimum-bias data and $2.0 \pm 2.0$ neutral current and $\nu_\tau$ events determined from simulation. An observed FD $\nu_\mu$ charge current rate of 33 events was seen, giving a clear signature of neutrino oscillations. 
Using these results the atmospheric neutrino oscillation parameters were measured to be $\sin^2(\theta_{23}) = 0.51 \pm 0.10$ and $\Delta m^2_{32}  = 2.37^{+0.16}_{-0.15} \times 10^{-3}$ eV$^2$ for the normal hierarchy and 
$\Delta m^2_{32}  = -2.40^{+0.14}_{-0.17} \times 10^{-3}$ eV$^2$ for the inverted hierarchy.  The energy spectrum of the observed events, along with the best-fit distribution to these events (with and without systematics) is shown in Figure~\ref{results}. The best-fit point along with the 68\% and 90\% contours in $\sin^2(\theta_{23})-\Delta m^2_{32}$ space for the normal hierarchy is also shown.

\section{Conclusion}
In conclusion, the extrapolation methods and systematic uncertainties associated with the analysis of the first data for the muon-neutrino disappearance analysis at NO$\nu$A have been described. This corresponded to $2.74\times10^{20}$ POT-equivalent collected between July 2013 and March 2015. 
This analysis is statistically limited and all systematic uncertainties are dominated by the absolute hadronic energy scale uncertainty.  
Fully quantifying the hadronic response will be essential for the next generation of results. 
With these results NO$\nu$A  has showcased its ability to produce world class physics and to be a leader in precision atmospheric neutrino oscillations measurements. With only 7.6\% of the nominal final statistics NO$\nu$A  is already competitive with the world limits.

\end{document}